\documentclass{article}
\usepackage[utf8]{inputenc} 
\usepackage[T1]{fontenc}    
\usepackage{hyperref}       
\usepackage{url}            
\usepackage{booktabs}       
\usepackage{amsfonts}       
\usepackage{nicefrac}       
\usepackage{microtype}      
\usepackage{xcolor}         
\usepackage{amsmath, amssymb, amsthm}
\newtheorem{theorem}{Theorem}[subsection]
\newtheorem{proposition}[theorem]{Proposition}

\usepackage{comment} 
\usepackage{todonotes}
\usepackage{caption}
\usepackage{subcaption}
\usepackage[linesnumbered,ruled]{algorithm2e}

\newcommand{\PP}{\mathbb{P}}

\newcommand{\getTitle}{A pragmatic policy learning approach to account for users' fatigue in repeated auctions}

\title{\getTitle}
\author{ Benjamin Heymann\thanks{Criteo AI Lab}, Rémi Chan-Renous\thanks{work done while interning at Criteo AI Lab},  Alexandre Gilotte\thanks{Criteo AI Lab}}
\begin{document}

\maketitle

\begin{abstract}

Online advertising banners are sold in real-time through auctions.
Typically,    the more banners a user is shown,  the smaller  the marginal value of the next banner for this user is. This fact can  be detected by 
 basic ML models, that can be used to  predict how  previously won auctions decrease the current opportunity value.  However, \textit{learning is not enough} to produce a bid that correctly accounts 
for how winning the current auction  impacts the future values.
Indeed, a policy that uses this prediction to maximize the expected payoff of the current auction could be dubbed  \textit{impatient} because such policy does not fully account for the repeated nature of the auctions.
Under this perspective, it seems that most bidders in the literature are \textit{impatient}.
Unsurprisingly, impatience induces a  cost.
We provide two empirical arguments for the importance of this cost of impatience. First,  an offline counterfactual analysis and, second,  a  notable business metrics improvement by mitigating the cost of impatience  with policy learning.

 \end{abstract}

\section{Introduction}

\subsection{RTB auctions}
Display advertising is a multibillion market, where online banners are typically sold through real-time bidding (RTB) platforms. 
While an internet user browses a website, an RTB platform hosts an online auction for each banner that could be displayed to the user. 
Because each auction occurs in real-time, some advertising intermediaries --- also known as \textit{demand-side platforms} (DSP) --- are in charge of bidding  on behalf of the advertisers. In what follows, we will prefer the term \textit{bidders} over DSP as it is more generic.

To do its job, a bidder assesses,  for each auction, the ad potential effect on  the advertisers' objectives using contextual features.
The typical industry practice consists  in multiplying the estimated probability of a conversion by the advertiser's value for a conversion. 
For instance, in the case of a campaign  optimized toward sales, letting $\alpha$ 
 be the amount the advertiser pays the DSP for each sale, $S$ and $D$ be the conversion event and the display event for a given ad, the expected payoff of the display would typically  be:
\begin{align}
\label{eq:traditionalformula}
    \underbrace{\alpha}_{\text{revenue per sales}} \times \underbrace{\PP(S|D)}_{\text{probability that a display will lead to a sale}}.
\end{align}
Once the display  value is computed according to a rule like Equation~\eqref{eq:traditionalformula}, the bidder outputs a bid which maximizes the immediate expected payoff\footnote{value - cost}, which is not equivalent to maximizing the long term payoff over the whole timeline.  For example, in a second price auction, this  means  bidding directly the value $ \alpha \times \PP(S|D)$. In general (when the auction is not second price)  the bid optimization   requires an estimation of the competition.
In  practice, the probability is estimated using logs of past display data, which contain features describing the displays and whether they lead to a conversion. 
An implementation of a logistic regression model to estimate the probability $\PP(S|D)$ is described in \cite{chapelle}.
\subsection{A short story to illustrate the cost of impatience}
Suppose you are offered the possibility to bid in a second price auction for a ticket that you can then exchange against a 100\$ bill. 
\textbf{How much should you bid?}
It is classical from auction theory (see for example~\cite{krishna2009auction}) that --- since the ticket value is 100\$--- your bid should be 100\$.

Now, suppose that  there will be two auctions: one in the morning and one in the afternoon. Also, you cannot exchange more than one ticket against a 100\$ bill, that is,  the afternoon ticket has no value if you won the morning auction.
Again, \textbf{"how much should you bid?"}
The answer is that you will be probably better off if you bid a bit less than 100\$ in the first auction. 
This phenomenon, that becomes stronger as the number of auctions during the day increases, is morally close to the \textbf{cost of impatience} we introduce next. 

\subsection{Connection with the users' fatigue}
\begin{figure}
    \centering
  \includegraphics[width=1.\linewidth]{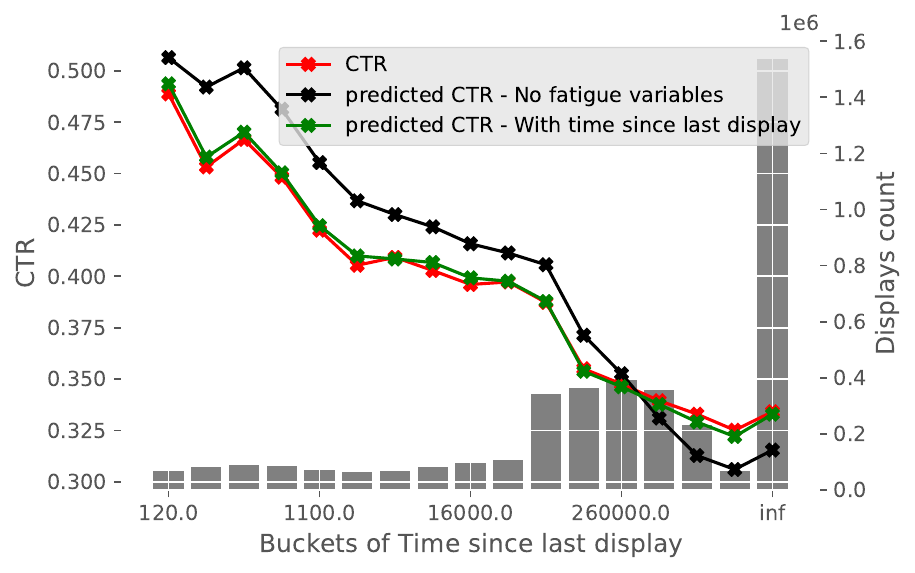}
  \caption{Empirical Click-Through-Rates (CTR) and predicted CTR computed on the \textit{Criteo Attribution Modeling for Bidding Dataset}~\cite{diemert2017pcb}.
  The green predictor uses a fatigue variable (the time since the last display), while the black one does not. 
  We observe that the predictor without fatigue variable tends to overpredict on recently exposed users. 
  }
  \label{fig:public_dataset}
\end{figure}

It is largely accepted that showing too many displays to the same user generates 'display fatigue' (see Figure~\ref{fig:public_dataset}), in other words the value of one additional display is decreasing with the number and/or frequency of the previous displays. A common practice in the industry is to use 'fatigue' variables in the prediction models, such as counters of past displays on the same user to improve the predictions.
However,  an optimal bidding policy  should also foresee that display fatigue  reduces the value of the next displays on the same user\footnote{Intuitively, if the user has several similar display opportunities shortly after, then the current opportunity should be valued less than its immediate expected reward since we could always try winning the display right after.

}.

More generically, the fact that the outcome of an auction has an impact on the future ones is largely ignored in display advertising. We propose to call this effect \textbf{the cost of impatience}. 
In \cite{ifa,mtc,diemert2017pcb}, the authors 
argue that,
when the value for an opportunity is impacted by past auctions outcome, then
learning to bid for repeated auctions shall be performed over the full timeline, and not at the auction level. However, they do not provide any operational tools to mitigate the cost of impatience. Taking a complementary perspective, ~~\cite{heymann2023repeatedbiddingdynamicvalue} presents a solution to the cost of impatience problem in a stylized setting. 

\subsection{Our contribution}
In this paper we  quantify the cost of impatience using counterfactual methods on real data from a major DSP.
We then introduce a method  inspired by policy learning~\cite{sutton2018reinforcement} to  mitigate the cost of impatience and test this solution on live traffic.
We were able to accurately predict the (online) effect of the changes at scale thanks to an (offline) Inverse Propensity Score estimator.

\section{Marginal analysis}

\subsection{IPS-based estimators}

We collected a dataset where we randomised a parameter $\Theta$ of the  bidder, by drawing for each user $i\in [1..n]$ a randomized value $\theta_i$ from a lognormal distribution.
We denote by $m_i$ a quantity of interest for a given user $i$\footnote{we insist that the index is on the user, not the auction} (such as the cost or the value generated by the won auctions) and we set $M(S)=\sum_{i\in S} m_i$ 
the aggregation of this quantity over
a set of user $S\subseteq [1..n]$.
The randomization allows the counterfactual estimation of outcomes
when the bidder changes the  policy, and draws $\Theta$ from another distribution. This counterfactual estimation relies on the  importance weighting estimators described in Proposition~\ref{eq:ips_estimator}.
More precisely, let $M(S,\alpha)$ be the expected value taken by $M(S)$ when  the lognormal random parameter $\Theta$ of each bidder is multiplied by $\alpha$, i.e.: 
$$ M(S,\alpha) :=\mathbb{E}_{ \Theta\sim\alpha\times Lognormal(\mu,\sigma)}\big[M(S)\big] $$

\begin{proposition}[Counterfactual estimator]
\label{eq:ips_estimator}
Let $S\subseteq [1..n]$ independent from $\Theta$ and  $\alpha > 0$, then

\begin{align}
  \widehat{M}(S,\alpha) :=  \sum_{i\in S} m_i \times \exp\left(\frac{2 \ln\left(\alpha\right) \left(\ln(\theta_i) - \mu\right) - \ln(\alpha)^2}{2 \sigma^2}\right)
\end{align}
is an unbiased estimator of $M(S,\alpha)$. 

\end{proposition}

\noindent Note  that the set of users $S$ must be defined \textit{independently} of the random factor $\Theta$ for the proposition to hold. To ensure that, we split the users in groups depending on their state (i.e. in our case their fatigue variable) at the \textit{beginning} of the data collection\footnote{More precisely, we looked at the state of the user at the time of the first bid-request after a random factor is drawn for this user.}.

\subsection{Linear approximation and marginal ROI}

For policy changes significantly larger than the standard deviation of the random exploration,
  estimator $\widehat{M}(S,\alpha)$ from Proposition~\ref{eq:ips_estimator} has a large variance~\cite{JMLR:v14:bottou13a}.
To avoid this variance, we used a linearized version of the importance weighting estimator. This linearised estimator trades of the variance for some bias, which we can expect to be reasonably small if the outcome is sufficiently regular with respect to the policy.
This linearised estimator is obtained by computing 
  the derivative   $\partial M(S,\alpha)/\partial \alpha$ at $\alpha=1$ 
  instead of directly using the IPS estimator $\widehat{M}  (S,\alpha)$.
This is possible by taking the derivative with respect to $\alpha$ at $\alpha=1$ in  Proposition~\ref{eq:ips_estimator}, as in the following proposition:

\begin{proposition}[Marginal counterfactual estimator]
\label{eq:estimator_dcost_dvalue}

Let $S\subseteq [1..n]$ independent from the $\theta_i$  set
\begin{align}
  \widehat{DM}(S) :=  \sum_{i\in S} m_i \times \frac{\ln(\theta_i) - \mu}{\sigma^2},
\end{align}
then $\widehat{DM}(S)$ is an unbiased estimator of the derivative of $\alpha\to M(S,\alpha)$ at $\alpha = 1$.

\end{proposition}

\begin{figure}
    \centering
  \includegraphics[width=.7\linewidth]{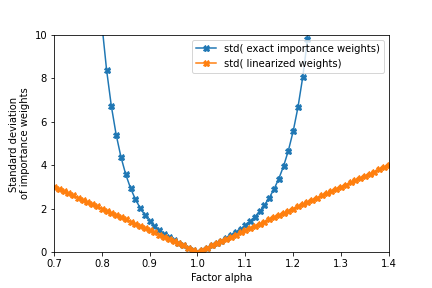}
  \caption{This figure shows the standard deviation of exact importance weights appearing in proposition \ref{eq:ips_estimator} as a function of the multiplicative factor $\alpha$; computed empirically from  lognormal samples. It grows exponentially with $\alpha-1$. On the other hand, the weights of the linearised estimator (in proposition \ref{eq:estimator_dcost_dvalue}) are directly proportional to $\alpha-1$, their standard deviation therefore grows linearly.}
  
  \label{fig:iw_std}
\end{figure}

\noindent Figure \ref{fig:iw_std} shows how the importance weights
standard deviation grows as $\alpha$ moves away from $1$.
As expected, standard deviation of exact importance weights quickly explodes, while standard deviation of the linear approximation grows linearly.

Given two quantities of interest $V$ and $C$, where $V$ is some form of value observed in the outcome, and $C$ is the money spent by  the bidder, we can now compute the \textbf{marginal ROI} on a given scope. 
Indeed, by the chain rule  and Proposition~\ref{eq:estimator_dcost_dvalue}, we have the following proposition. 
\begin{proposition}[marginal ROI]
Let $S\subseteq [1..n]$ independent from $\theta_i$, set
\begin{align}
\label{eq:estimator}
    mROI(S) = \frac{\widehat{DV}(S)}{\widehat{DC}(S)},
\end{align}
then $mROI(S)$ is a consistent estimator of the marginal ROI on $S$
\end{proposition}
\noindent The marginal ROI $mROI(S)$ can be interpreted as the incremental value $\Delta V$ the bidder gets from $S$ by spending one additional unit of money on  $S$.

\subsection{Maximising the value at constant cost}

We assume the bidder's task is to maximize the value $V_\pi$ for a given budget $B$:
\begin{align}
    \max_{\pi} V_\pi\\
    \mbox{s.t.}\quad  C_\pi\leq B
\end{align}
where $V_\pi$ and $C_\pi$ are the overall value and spend generated by the bidder's policy.
Clearly, if  there exists two set of users $S_1$ and $S_2$ such that
\begin{align}
    S_1\cap S_2= \emptyset\quad \mbox{and}\\
    mROI(S_1)\leq mROI(S_2),
\end{align}
then the bidder can improve the criterion by rebalancing some budget from $S_1$ to $S_2$.
Said otherwise, if two groups of users have different marginal ROIs, the value can be increased by spending more on the users with a high ROI and less on users with a smaller ROI.
More generally, suppose we have a clustering 
 $S_1, S_2\ldots S_k$  of $[1..n]$ such that 
\begin{align}
    mROI(S_i)\leq mROI(S_{i+1}),\quad \forall i \in [1,k-1].
\end{align}
\noindent A naive idea could be to increase \footnote{Assuming here that cost and value are increasing with $\alpha$} away from 1 the factor $\alpha$ to a high value on $S_k$ and to decrease it everywhere else. This is obviously not a wise idea: the $mROI$ only tells what happens for small perturbations of the parameter $\Theta$. In practice the marginal ROI is usually a decreasing function of the spend, so that at some point $S_k$ won't be the best cluster to spend on.  
We thus propose a straightforward decision rule that consists in capping $\alpha$ at a reasonable level around 1 \footnote{In our experiments we capped the factor $\alpha$ to the $[0.8,1.2] $ interval } on the different clusters, and then to choose for each cluster $S$ a value $\alpha_S$ within this range such that, according to the linearised estimator, the total cost variation $\sum\limits_S \sum\limits_{i \in S } (\alpha_S - 1) \cdot c_i \cdot \frac{ log(\theta_i)-\mu }{\sigma^2}  $ is 0 and the total value is maximised.

\section{Experiments}
\subsection{Analytics}
\paragraph{Proxy for the reward} The \textit{value} in the bidding system is defined as the number of conversion events matched to the displays, multiplied by a predefined value per conversion. However, these conversion events are scarce, which means that the estimator of the policy value have a significant variance. To reduce this variance, we replaced the actual count of conversions by the expected number of conversions, computed with the  prediction model already used by our baseline bidder. \footnote{This idea of replacing actual reward by predicted reward is a common used in RL in 'actor-critic' algorithms.}

\begin{figure}[htp!]
    \centering
\begin{subfigure}{0.49\textwidth}
  \includegraphics[width=1.1\linewidth]{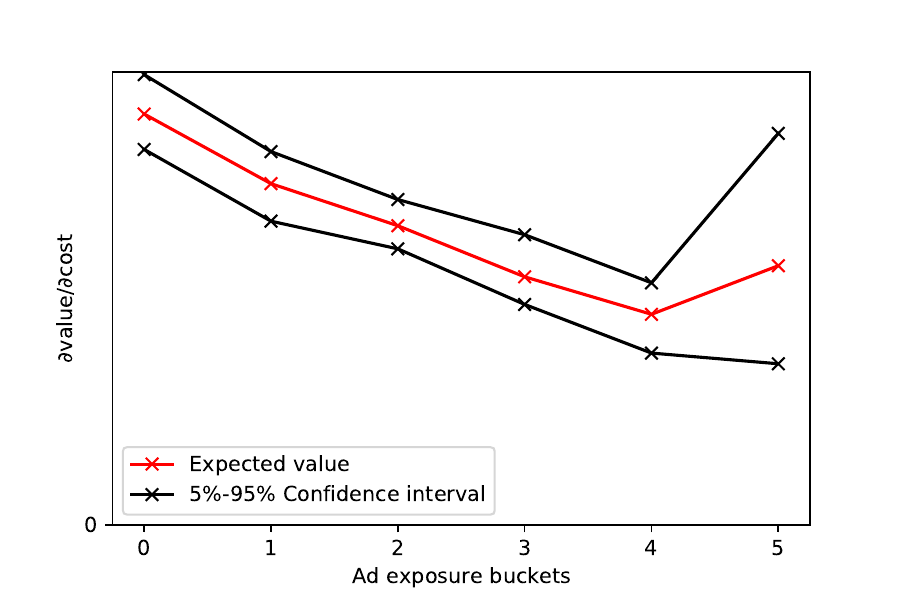}
  \caption{02/2022}
  \label{subfig:02/2022}
\end{subfigure}\hfil
\begin{subfigure}{0.49\textwidth}
  \includegraphics[width=1.1\linewidth]{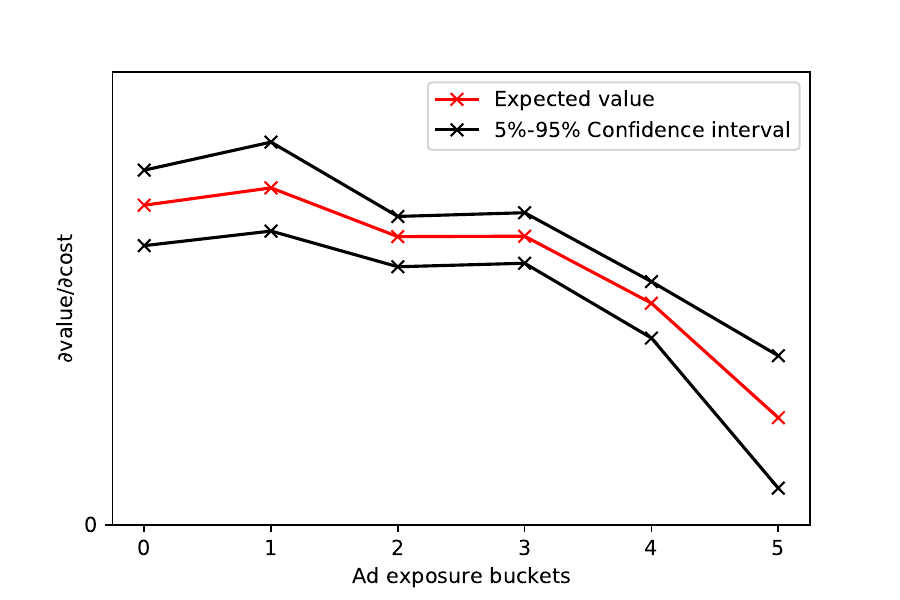}
  \caption{03/2022}
  \label{subfig:04/2022}
\end{subfigure}\hfil
\caption{The plots \ref{subfig:02/2022} and \ref{subfig:04/2022} represent the confidence intervals of the marginal ROI computed for different levels of ad exposure. Each plot uses sampled data from one month. We processed only the data for which we have access to the ad exposure. The bucket 5 corresponds to extreme values for which we have fewer samples, resulting in larger confidence intervals. Intuitively  --- because winning an auction decreases the values of the future auctions --- the bidder should decrease the bid when  it forecasts to receive many bidding opportunities.}
\label{fig:roi}
\end{figure}

\paragraph{Observed marginals}
We computed the marginal ROI estimator from \eqref{eq:estimator} as a function of the user's ad exposure\footnote{with the definition of ad exposure we found the most relevant} 
on several randomised datasets coming from different time periods. We report some   results in Figure \ref{fig:roi}. We note that the marginal return on investment  is decreasing with the ad exposure. In a nutshell, increasing the spend  by one unit for users with little ad exposure results in a steeper increase in the value  than doing the same for users more exposed to the ad, 
and this is consistent with our initial intuition.

\subsection{ Pre-A/B test offline estimation of the resulting policy }
\begin{figure}
    \centering
    \includegraphics[width=1.\linewidth]{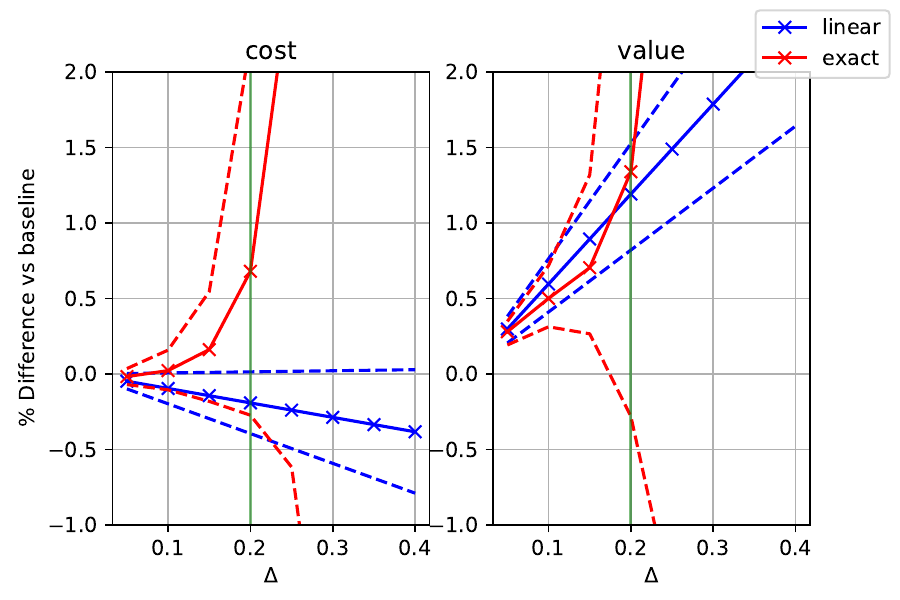}
    \caption{Pre-A/B test offline estimation of the resulting policy, $\Delta$ on the x axis is the maximum amplitude of the change of parameter.The green vertical line corresponds to the amplitude of change tested during the online A/B test.
    We see that the linearization drastically reduce the confidence intervals (dotted lines).}
    \label{fig:offline_results}
\end{figure}
We did a counterfactual estimation of the value and cost generated by some changes of policy using the linear and exact IPS formulas. 
We used three months of data to build the policies and 2 month of data to estimate the resulting performances. 
The results are displayed in Figure~\ref{fig:offline_results}. 
We observe that it is likely that the setup will decrease the cost and increase the value generated.

\subsection{Live experiments}
We tested the modified bidding module by doing a random split of the traffic, and assigning the new module to one of the user group. 
In the online experiment however, we used the \emph{current state} of the user fatigue at each display opportunity to retrieve the  factor $\alpha$. This policy can thus change the  factor on a user who viewed many displays: this is not strictly identical to the policy we tested offline, where the user was assigned to a fatigue bucket once at the beginning of the data collection process.

There are two reasons for this discrepancy. One is simplicity, as  it is in practice much easier to access the current user fatigue variable than to retrieve its value at the beginning of the A/B test.  The other reason is that it is intuitively much more relevant to use the current state. Actually, the reason why our offline policy had fixed bid factors per users was because the randomised data we collected only contained a randomisation at the user level, and thus did not allow simulating richer policies which would change the bid factor during a user sequence. We thus view the offline policy with fixed factors as an easy-to-simulate approximation of the intuitively better dynamic policy which we A/B tested.
In accordance with our offline estimations, the A/B test was positive, producing  an increase of around $0.7\%$ in  value and a decrease of around $- 1\% $ in cost.  
In a system as mature as the one on which this test was performed, such an outcome is a great achievement.  

\section{Wrapping-up}

We use this section to summarize the ideas we used in this experimental study. 

\begin{enumerate}
    \item get randomized data on the parameter to be fine tuned $\to$ unlock counterfactual analysis
    \item rely on the prediction of reward rather than the reward $\to$  reduce the variance in the decision problem
    \item do a first order approximation of the IPS $\to$  tame the variance of the IPS estimator
    \item  compute the marginal ROI on the different clusters $\to$ decide where to reallocate the budget
    \item check the effect of the new policy with IPS and bootstraps   $\to$ predict the online effect
    \item test online
\end{enumerate}
It is notable that this recipe  is  close  to a manual step of reinforce. 
We believe this approach to be generic and that it could be adapted to other use cases. 
As a consequence, further work includes extending the design to allow for automated multi-step learning.

\newcommand{\etalchar}[1]{$^{#1}$}

\end{document}